\documentclass[aps,preprint,superscriptaddress, pra]{revtex4-1}
\usepackage{lipsum}
\usepackage{epsfig}
\usepackage{amssymb,amsmath,amsthm}
\usepackage{color}
\usepackage[colorlinks=true,linkcolor=blue,citecolor=blue]{hyperref}

\begin{document}


\title{Drift instabilities in localised Faraday patterns}




\author{\copyright Juan F. Mar\'in}
\email[]{juan.marin.m@usach.cl}
\affiliation{Departamento de F\'isica, Universidad de Santiago de Chile, Usach, Av. V\'ictor Jara 3493, Estaci\'on Central, Santiago, Chile.}

\author{Rafael Riveros \'Avila}
\affiliation{Instituto de F\'isica, Pontificia Universidad Cat\'olica de Valpara\'iso, Casilla 4059, Chile}

\author{Saliya Coulibaly}
\affiliation{Universit\'e de Lille, CNRS, UMR 8523, Laboratoire de Physique des Lasers Atomes et Mol\'ecules, F-59000 Lille, France}

\author{Majid Taki}
\affiliation{Universit\'e de Lille, CNRS, UMR 8523, Laboratoire de Physique des Lasers Atomes et Mol\'ecules, F-59000 Lille, France}

\author{M\'onica A. Garc\'ia-\~Nustes}
\affiliation{Instituto de F\'isica, Pontificia Universidad Cat\'olica de Valpara\'iso, Casilla 4059, Chile}


\date{\today}

\begin{abstract}
Nature is intrinsically heterogeneous, and remarkable phenomena can only be observed in the presence of intrinsically nonlinear heterogeneities. Spontaneous pattern formation in nature has fascinated humankind for centuries, and the understanding of the underlying symmetry-breaking instabilities has been of longstanding scientific interest. In this article, we provide theoretical and experimental evidence that heterogeneities can generate convection (drift instabilities) in the amplitude of localised patterns.  We derive a minimal theoretical model describing the growth of localised Faraday patterns under heterogeneous parametric drive, unveiling the presence of symmetry-breaking nonlinear gradients. The model reveals new dynamics in the phase of the underlying patterns, exhibiting convective instabilities when the system crosses a secondary bifurcation point. We discuss the impact of our results in the understanding of convective instabilities induced by heterogeneities in generic nonlinear extended systems far from equilibrium.
\end{abstract}

\pacs{
05.45.Yv 
05.45.-a, 
89.75.Kd  
}

\maketitle



\section{Introduction\label{sec:introduction}}

\begin{figure*}
\begin{centering}
    \begin{minipage}{0cm}
    \scalebox{0.35}{\includegraphics{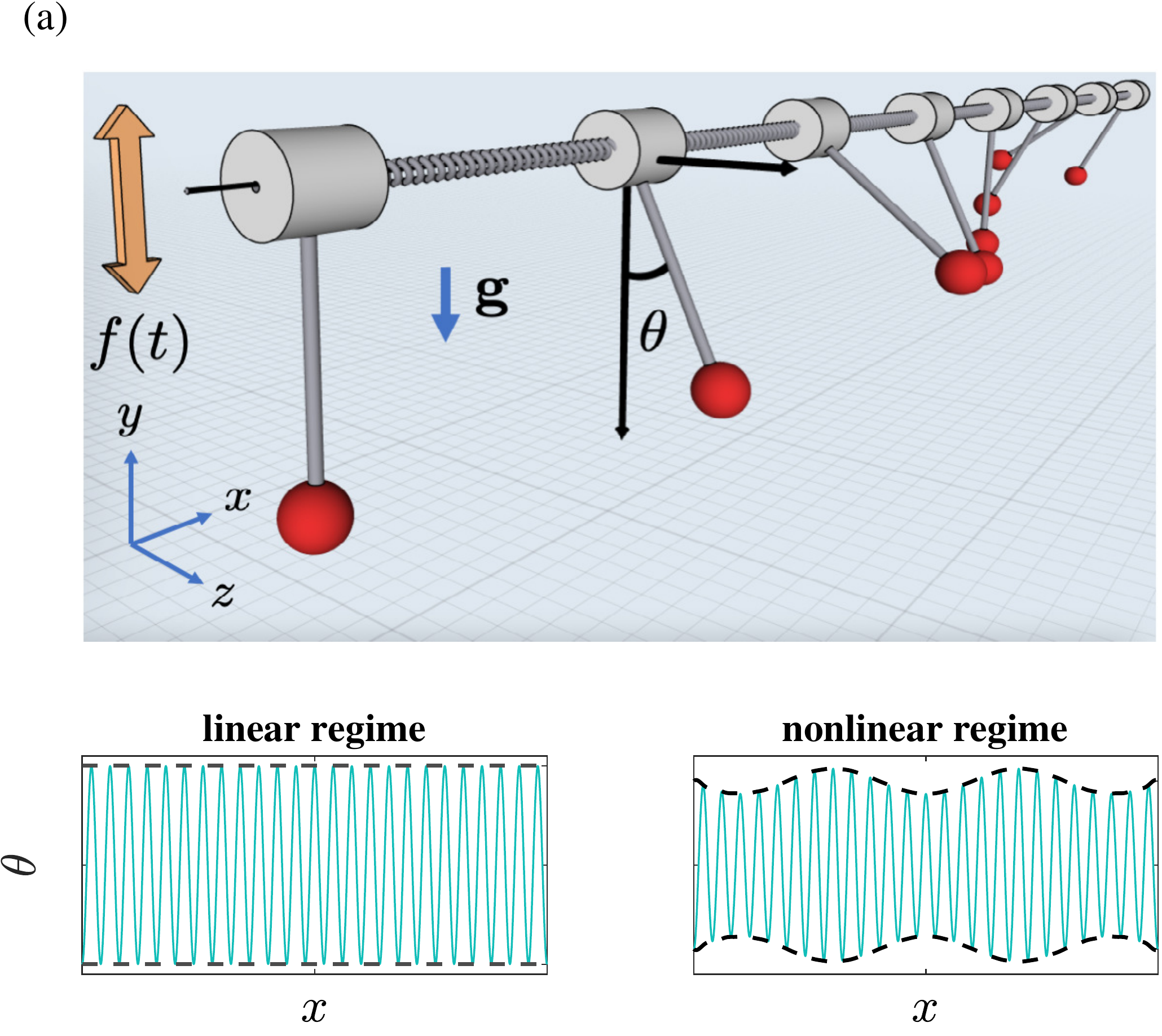}}
    \end{minipage}
    \hfill
    \   \ 
    \hfill
    \begin{minipage}{9.1cm}
    \scalebox{0.38}{\includegraphics{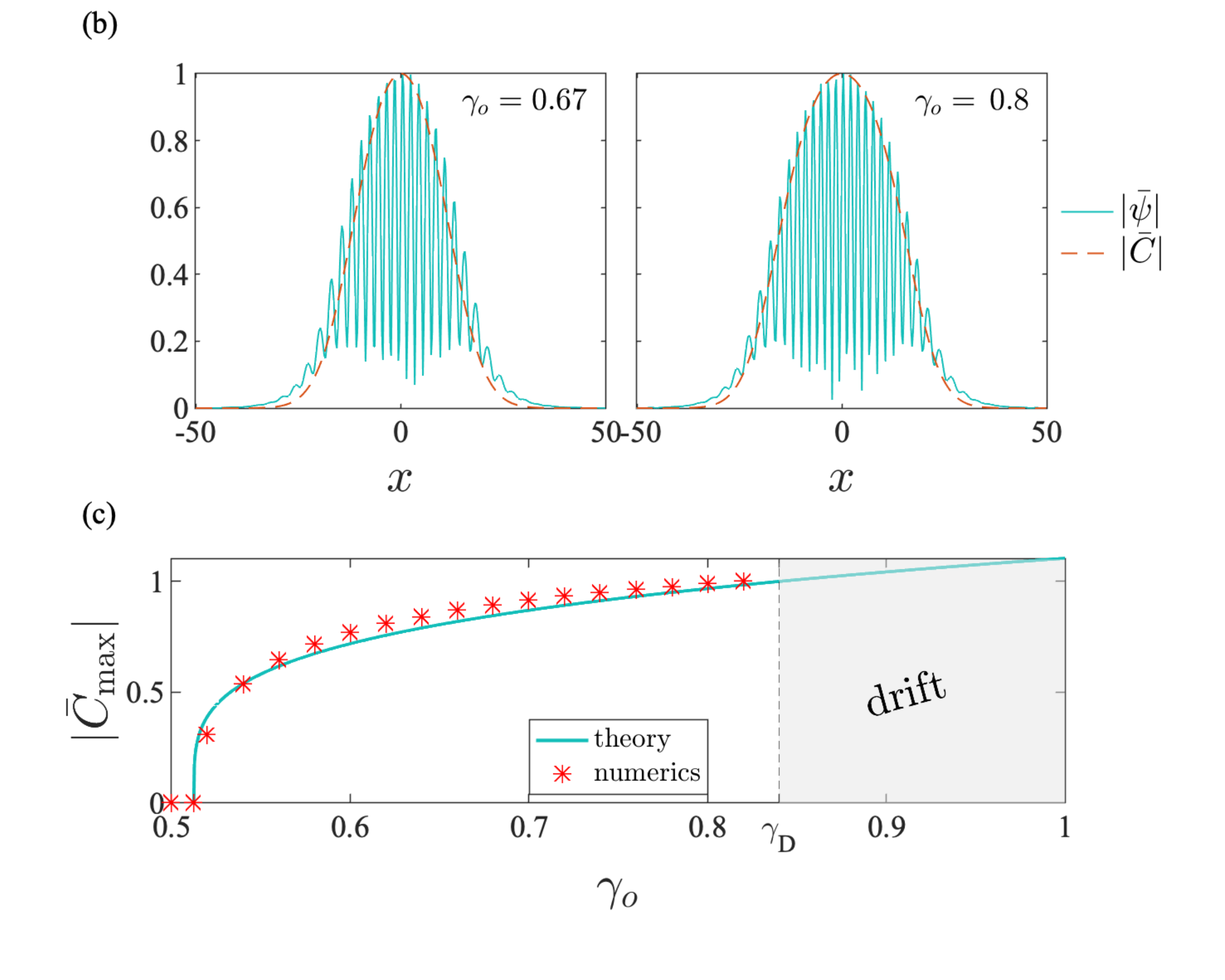}}
    \end{minipage}
    \par\end{centering}
    \caption{\textbf{(a)} Schematic of a system of coupled pendula under parametrical drive as a model to study pattern formation phenomena. The envelope of linear plane waves becomes modulated as nonlinear harmonics grows in time, generating a pattern with a characteristic wavelength and frequency. \textbf{(b)} Normalised amplitudes $\vert\bar\psi\vert$ and $\vert\bar C\vert$ obtained numerically from Eq.~\eqref{Eq:01} and Eq.~\eqref{Eq:04}, respectively, with $\mu=0.45$, $\nu=1$, $\alpha=1$, and $\sigma_i=16$. \textbf{(c)} Normalised maximum amplitude of the saturated Faraday wave, $|\bar{C}_{\max}|$, as a function of $\gamma_o$, according to Eq.~\eqref{Eq:05} (solid line) and the numerical solutions of the pdnlS Eq.~\eqref{Eq:01} (stars). Drift is observed above a critical pump $\gamma_D\simeq0.837$ indicated with a vertical dashed line.} 
    \label{fig:01}
\end{figure*}

Patterns are ubiquitous in non-equilibrium systems \cite{Cross2009}. They appear in very different systems, such as fluids, chemical reactions, magnetic textures, nonlinear optics, and granular layers \cite{Pismen2006, Miles1984, Malomed2009, perinet2009, Clerc2010}. A paradigmatic toy-model for the study of pattern formation in extended non-equilibrium systems is the wave-envelope description in the continuum limit of a chain of coupled nonlinear dissipative oscillators, such as the torsion pendula depicted in Fig.~\ref{fig:01}(a). Besides its simplicity, such system has proven to give profound insights into the generic phenomenon of pattern formation. A vertically-oscillating drive $f(t)=f_0\cos\Omega t$ forcing the common axis of the pendula has the effect of modulating the effective gravity-- a parameter of the system --, and when the amplitude of such driving is above certain threshold value the system undergoes the celebrated parametrical instability \cite{LandauMechanics}. Above the threshold of instability, the homogeneous envelope of a plane wave turns unstable and become modulated due to the development of weakly nonlinear harmonics-- see the insets below Fig.~\ref{fig:01}(a). One observes in the envelope the spontaneous formation of a regular pattern oscillating with a well defined wavelength and a sub-harmonic response, i. e. oscillating with frequency $\omega=\Omega/2$. These patterns are known as \textit{Faraday patterns}, and were first reported by Faraday in his seminal experimental work on vertically vibrated fluids and granular layers \cite{Faraday1831-II, Faraday1831}. Just above the threshold of instability, the amplitude of the pattern grows with a slow characteristic time until saturation. Saturation is provided by a double dynamical compensation: one between nonlinearity and dispersion, and another between dissipation and the parametric drive. In such an envelope description, taking the continuum limit in the weakly nonlinear analysis of the system of coupled pendula, one obtains that the amplitude is governed by the following parametrically driven nonlinear Schr\"odinger (pdnlS) equation\cite{Alexeeva2000},
\begin{equation}
\label{Eq:01}
\partial_t\psi=-(\mu+i\nu)\psi-i\alpha\partial_{xx}\psi-i|\psi|^2\psi+\gamma(x)\psi^*,
\end{equation}
which is written in dimensionless form. In this model, pattern dynamics is described through a complex order parameter $\psi$ that depends on space and time. Parameter $\mu$ is the dissipation, $\alpha$ is the diffraction length, $\nu:=\Omega/2-\omega_o$ is the detuning between the sub-harmonic frequency of the generated Faraday wave and the natural frequency $\omega_o$ of the system, and $\gamma_o:=f_o/4\omega_o$ is the normalised amplitude of the drive (pump).

If $\gamma=\gamma_o=\hbox{constant}$, spatial instabilities occur if $\gamma>\gamma^h_c:=\mu$~\cite{Miles1984}. In such scenario, the instability spontaneously grows into a pattern formation for positive values of $\nu$ through a supercritical bifurcation \cite{Coullet1994, Clerc2009} with leading eigenmodes of wave-number $\kappa_c=\pm\sqrt{\nu/\alpha}$. The effect of a heterogeneous pump in this system has been recently studied \cite{Urra2019} by introducing the space-dependent Gaussian profile $\gamma(x)=\gamma_o\exp\left({-x^2/2\sigma_i^2}\right)$, where $\gamma_o$ is the pump strength and $\sigma_i$ its extension. The linear stability analysis under a slightly non-uniform Gaussian drive, i.e. $\sigma_i \ll \sqrt{\alpha}$, shows that the envelope $C_0$ of the pattern has a discrete response given by the Gauss-Hermite polynomials \cite{Urra2019}. Urra et al.~\cite{Urra2019} showed that the threshold of the instability of the $m$-th Gauss-Hermite mode is given by
\begin{equation}
\label{Eq:02}
\gamma_o^{(m)}=\mu+(2m+1)\frac{1}{\sigma_i}\sqrt{\frac{\nu}{\alpha}}.
\end{equation}
Thus, localised Faraday patterns appear if $\gamma_o>\gamma_o^{(0)}$. Near the fundamental threshold of instability, Faraday patterns are rather regular and highly ordered both in the real and imaginary parts of $\psi$ as well as on its phase, $\arg(\psi)$. The amplitude of the pattern increases as we increase the value of $\gamma_o$. However, at some point in the interval $(\gamma_o^{(2)},\gamma_o^{(3)})$, a secondary instability emerges and pattern dynamics become complex both in amplitude and phase. In many cases, a secondary instability is a precursor to spatiotemporal chaos and transition to turbulence \cite{Zhang1995, Schmid2001}.

In this article, we use the normal form theory to formally derive an amplitude equation for localised Faraday patterns induced by heterogeneous parametric drive. As will be shown, the evolution of the instabilities at the first order of nonlinearity is described by a quintic Complex Ginzburg-Landau equation with Webber-like and nonlocal self-phase modulation terms in the form of nonlinear gradients. Such nonlinear gradients triggers convection, i.e.  \emph{drift instabilities}, through a spontaneous nonlinear symmetry breaking above a secondary bifurcation point. Importantly, such drift is entirely induced by the heterogeneity of the pump and cannot be observed under homogeneous drive. Theoretical conditions for drift appearance are reproduced in a fluid experiment where Faraday waves are induced by a localised bottom motion in a water cell, providing experimental confirmation of the predicted drift.

\section{Results\label{sec:introduction}\label{Sec:Results}}

\subsection{Theoretical prediction of convection in localised Faraday waves}

We have characterised the dynamics of localised Faraday waves near the fundamental threshold of instability through Eq.~\eqref{Eq:01} using the method of normal forms [see Section \ref{Section:Methods}: Methods]. It is a widely used formalism in which a nonlinear system takes the ``simplest''-- or so called \emph{normal} -- form preserving the essential features of the system, such as its behaviour near a bifurcation point~\cite{Haragus2010}. To characterise the dynamics close to the spatial instability, we introduce a bifurcation parameter $\delta:=\gamma_o-\gamma_o^{(0)}$ and a small dimensionless parameter $\epsilon:=\sqrt{\alpha}/\sigma_i\ll1$ characterising the length of the injection zone. With the condition $\epsilon\ll1$ we are assuming that the injection zone is large compared with the diffraction length, i.e. a weakly heterogeneous system. We study the behaviour of solutions around a Faraday wave with critical wavenumber $\kappa_c$ introducing a modulating complex amplitude function $C(x,t)$ -- an \emph{order parameter} --, i.e.
\begin{equation}
\label{Eq:03}
\psi(x,t)=C(x, t)e^{i\kappa_c x} + c.c.+ h.o.h.,
\end{equation}
where $h.o.h$ denotes the high-order harmonics. Below the bifurcation point, $\gamma_o<\gamma_o^{(0)}$, one obtains the stable homogeneous solution $C=0$ (disordered phase). Such homogeneous solution becomes unstable just above the bifurcation point, and $C(x,t)$ has its own spatiotemporal dynamics modulating the underlying Faraday pattern (self-organised phase). At the leading order of non-linearity, one obtains the following governing equation for the amplitude,
\begin{equation}
\label{Eq:04}
\partial_tC=\frac{2\nu}{\alpha\mu}\partial_{xx}C+\left(\delta-\frac{\mu}{2\sigma_i^2}x^2\right)C-i\frac{2}{\mu}\sqrt{\frac{\nu}{\alpha}}\left[3\partial_x(|C|^2C)-2C\partial_x(|C|^2)\right]-\frac{9}{2\mu}|C|^4C.
\end{equation}

For non-extended systems, the homogeneous limit of Eq.~\eqref{Eq:04} turns into the well-known amplitude equation previously derived by Coullet \textit{et al.} \cite{Coullet1994}. The first two terms in the right-hand side of Eq.~\eqref{Eq:04} are Weber-like terms that give the known linear limit of the heterogeneous system \cite{Urra2019}. At the modulational instability, the amplitude of the unstable Gauss-Hermite modes is saturated by the quintic nonlinearity given by the last term in Eq.~\eqref{Eq:04}. Importantly, the terms proportional to $\partial_x(|C|^2C)$ and $C\partial_x(|C|^2)$ in Eq.~\eqref{Eq:04} are new non-local terms unveiling \textit{self-phase modulation} (SPM) effects in the slowly evolving modes of the localised Faraday patterns. Such terms are commonly introduced in the nonlinear Schr\"odinger equation to account for self-steepening effects and self-frequency shifting via stimulated Raman scattering \cite{Kumar1990, Porsezian1996, Gedalin1997, Nakkeeran1998, Liu2009}. To our knowledge, this is the first time a complex quintic Ginzburg-Landau equation with Webber-like terms has been derived with these non-local terms appearing naturally from the weakly nonlinear analysis. The appearance of such terms lead us to discover absolute and drift instabilities in the localised Faraday waves, as we show below.

In Fig.~\ref{fig:01}(b) and Fig.~\ref{fig:01}(c) we compare the predictions of our amplitude equation and the pdnlS equation. There is a remarkable agreement between both models near the fundamental threshold of instability. In Fig.~\ref{fig:01}(b), we solve numerically Eqs.~\eqref{Eq:01} and \eqref{Eq:04} with no-flux boundary conditions. The initial conditions are given by small-amplitude random distributions of values in space for the real and imaginary parts of $\psi$ and $C$.  Figure~\ref{fig:01}(b) confirms that the extension $\sigma$ and the envelope profile of the Faraday pattern are well described by the amplitude equation. As one would expect, predictions from the amplitude equation show some mismatch from the outcome of the pdnlS equation away from the fundamental threshold. Figure~\ref{fig:01}(c) demonstrate that the amplitude equation~\eqref{Eq:04} well describes the nature of the bifurcation of the Faraday pattern, as we will see later.

\begin{figure*}
\begin{centering}
    \scalebox{0.184}{\includegraphics{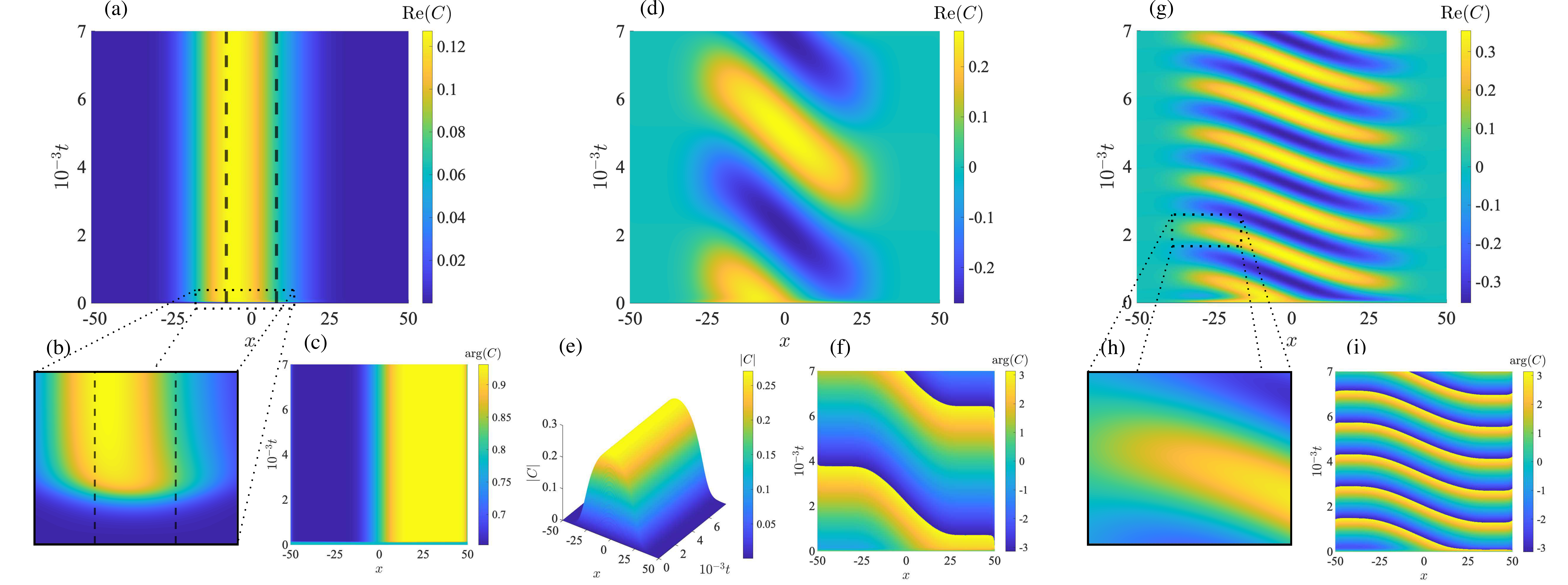}}
    \par\end{centering}
    \caption{Drift instabilities in the amplitude equation \eqref{Eq:04}. Numerical results were obtained with $\mu=0.45$, $\nu=1$, $\alpha=1$, and $\sigma_i=16$. \textbf{(a, b, c)} A wave exhibiting absolute stability for $\gamma_o=0.80$. \textbf{(d, e, f)} Convective instability for $\gamma_o=1.2$. \textbf{(g, h, i)} A strongly nonlinear convective instability for  $\gamma_o=1.62$.}
    \label{fig:02}
\end{figure*}

The SPM terms in Eq.~\eqref{Eq:04} are induced by nonlinear gradients, which reveal underlying complex dynamics in the phase. The gradient terms break the transverse reflection symmetry ($x\to-x$) and dramatically affect the pattern selection in the system \cite{Louvergneaux2004, Coulibaly2008}. Thus, the nonlinear gradients in Eq.~\eqref{Eq:04} give a \emph{nonlinear symmetry breaking} where convection is characterised by an amplitude-dependence in the group velocity. In Fig.~\ref{fig:02}, we demonstrate drift instabilities in the system, which has been obtained from the numerical simulation of Eq.~\eqref{Eq:04} for the given values of parameters.

In Fig.~\ref{fig:02}(a), we show a scenario where amplification of the wave-amplitude dominates, and the system exhibits absolute stability. In the absence of SPM terms, the localised Faraday wave spontaneously grows centred around the Gaussian pump with width $\sigma_w=(\sqrt{\nu}\sigma_i/\mu)^{1/2}$~\cite{Urra2019}. In Fig.~\ref{fig:02}(a), we indicate with dashed lines the with $\sigma_w$ centred at the Gaussian pump for reference, evidencing a clear shift in $\mbox{Re}(C)$ due to the SPM terms. Figure~\ref{fig:02}(b) shows a zoom-in at the beginning of the simulation, showing that there is indeed a transient drift towards the left as the pattern grows. However, convection is too weak to overcome saturation, and the system rapidly becomes absolutely stable. Although the imaginary part of $C$ exhibits a similar shift as the real part, the amplitude $|C|$ exhibit no shift and the phase shows no spatiotemporal dynamics in this case, as shown in Fig.~\ref{fig:02}(c).

Increasing the value of $\gamma_o$, we found for some value $\gamma_{\text{D}}\in(\gamma_o^{(2)},\gamma_o^{(3)})$ that the propagation overcomes amplification and the system exhibits a nonlinear transition to convective instability. This scenario is shown in Fig.~\ref{fig:02}(d), evidencing a drift in the real part of the envelope of Faraday waves. The real and imaginary parts of $C$ exhibit the same drift, excepting a phase factor. However, the amplitude $|C|$ of the wave has no drift, as we show in Fig.~\ref{fig:02}(e). Thus, convection in the system is entirely due to the phase dynamics, as evidenced in Fig.~\ref{fig:02}(f).

A further increase in $\gamma_o$ increases the drift speed, as shown in Fig.~\ref{fig:02}(g). The drifting wave also experiences a change in both frequency and wavelength. These observations confirm the nonlinear nature of the drift instability. Notice that the drifting pattern becomes non-propagating near the tails of the Gaussian pump, as highlighted in the zoom-in of Fig.~\ref{fig:02}(h). This pinning effect stems from the spatial heterogeneity of the pump, which imposes a smooth amplitude variation at the tails of the Gaussian inducing a boundary condition-like. Such kind of boundary conditions induces spatial variations for the pattern envelope that can become comparable to its modulation amplitude. This results in the coupling between the slow scale of the envelope to the fast scale of the modulation of the underlying Faraday pattern, leading to the pinning of the drifting wave \cite{Clerc2012}. Moreover, the phase dynamics exhibit travelling waves inside the injection zone defined by the Gaussian parametrical drive, as evidenced in figures~\ref{fig:02}(f) and \ref{fig:02}(i). The mean speed of such travelling waves decreases as $\gamma_o$ decreases.

\subsection{Drift in the model system}

\begin{figure}
\begin{centering}
   \scalebox{0.38}{\includegraphics{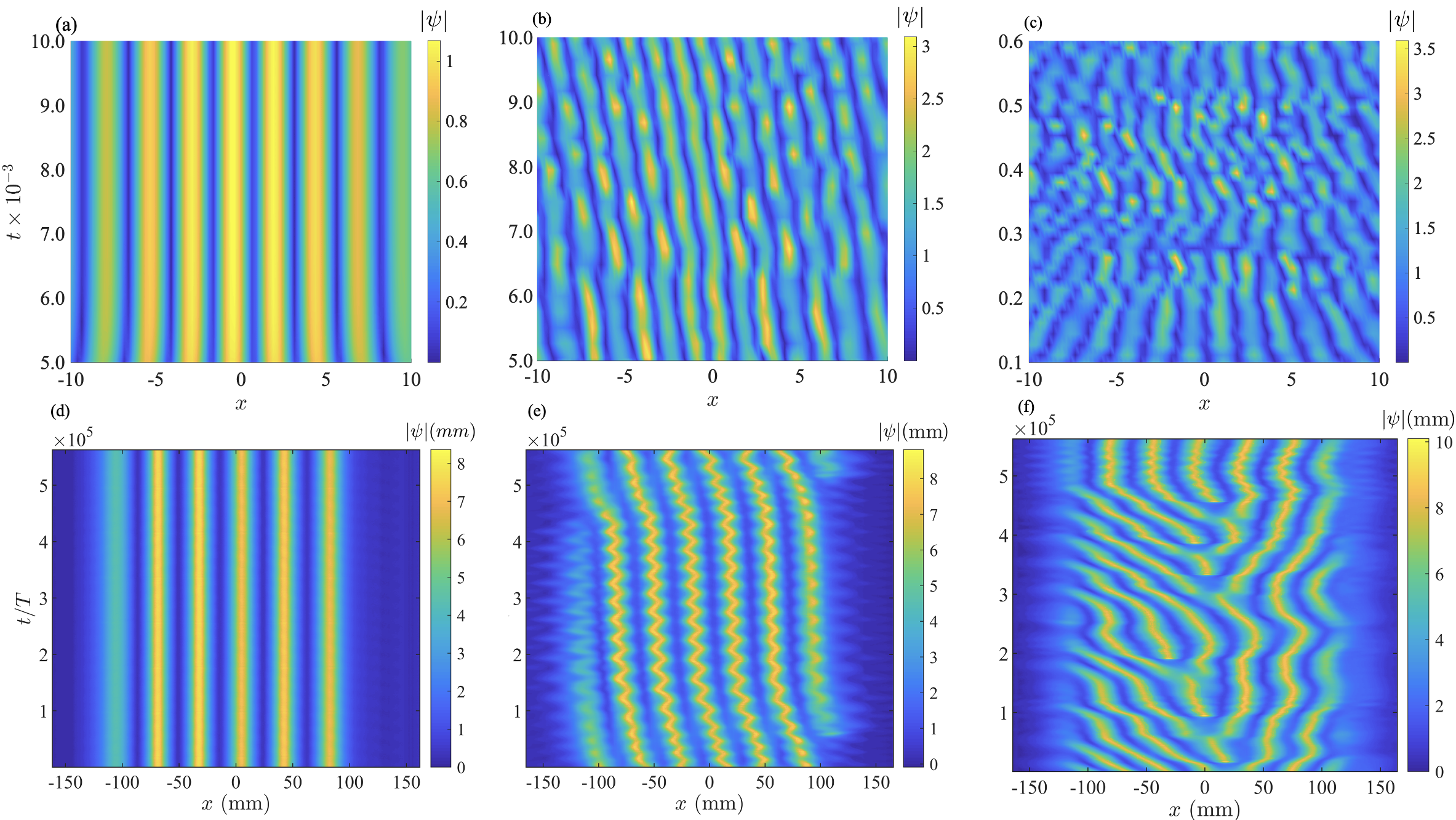}}
    \par\end{centering}
    \caption{\textbf{(Upper panel)} Drift instability found in the numerical simulations of the pdnlS equation \eqref{Eq:01} with $\mu=0.45$, $\alpha=1$, $\nu=1.0$, and $\sigma_i=16$ for \textbf{(a)} $\gamma_o=0.55$, \textbf{(b)} $\gamma_o=0.8375$, and \textbf{(c)} $\gamma_o=0.9000$.
    \textbf{(Lower panel)} Experimental evidence of the drift instability found in a fluid subject to vertical and localised vibrations. The amplitude of the vertical acceleration is \textbf{(d)} $\Gamma_0=0.56$, \textbf{(e)} $\Gamma_0=0.70$, and \textbf{(f)} $\Gamma_0=0.84$. The detected height of the surface of the fluid is shown in the colour scale. } 
    \label{fig:03}
\end{figure}

Given that the amplitude equation \eqref{Eq:04} predicts drift above a secondary threshold of instability, we consider again the pdnlS model, Eq.~\eqref{Eq:01}, to find out if it exhibits such drift in the underlying Faraday patterns. In the upper panel of Fig.~\ref{fig:03}, we show the results from the numerical simulation of the pdnlS model for the given values of the parameters.

In Fig.~\ref{fig:03}(a), we have fixed $\gamma_o$ slightly below the secondary instability onset $\gamma_D$ and the localised pattern is absolutely stable, as expected. Notice that the maximum of the wave amplitude is shifted towards the left of the origin, in accordance with the behaviour observed in Fig.~\ref{fig:02}(a). Remarkably, if $\gamma_o$ is slightly above the threshold $\gamma_D$, the unidirectional drift becomes evident, as shown in Fig.~\ref{fig:03}(b). These results evidence the nonlinear symmetry breaking in the system and confirm the predictions from our normal form \eqref{Eq:04}. Further increasing the pump $\gamma_o$ above the threshold of convective instability, the system evolves towards a more complex convective state with a superposition of oscillations (suggesting an underlying Hopf instability) and drift towards both directions as shown in Fig.~\ref{fig:03}(c). Remarkably, as we will discuss later in this article, we have found in a fluid experiment the same drift within the range of parameters predicted by our normal form, as shown in Fig.~\ref{fig:03}(d-f).

To gain further insight into the dynamical behaviour of the drifting localised patterns, we derive an explicit expression for the pattern speed as a function of parameter $\gamma_o$ following a similar procedure as in Ref.~\cite{Clerc2012}. We write our amplitude equation Eq.~\eqref{Eq:04} in the polar form by setting $C=\rho\exp(i\theta)$. After removing the phase factor, the imaginary part reads
\begin{equation}
\label{Eq:PolarForm}
\partial_{\tau}\theta=4\partial_{XX}\theta+\frac{8}{\rho}\partial_X\rho\partial_X\theta-20\rho^2\partial_X\rho,
\end{equation}
with $X:=x\sqrt{\nu/\alpha}$ and $\tau:=2\mu t$. As depicted in Figs.~\ref{fig:02}(f) and \ref{fig:02}(i), the phase $\theta(x,t)=\arg(C)$ is composed of two different superimposed dynamical behaviours: (i) a travelling wave with some phase speed $v=\omega/k$ and (ii) a monotonically increasing one modulo $2\pi$, such that $\arg(C)\in[-\pi,\,\pi]$ (following the conventional restriction of the inverse trigonometric functions to the principal branch). Based on this observation, we propose the ansatz $\theta(X,\,\tau)=\theta_v(X-{\color{blue}v}\tau)+\chi(\tau)\mod(2\pi)-\pi$, where $\theta_v$ and $\chi$ account for the travelling and the increasing dynamics, respectively. Next, let us take the space and time average $\langle\theta(X,\tau)\rangle_X=\sigma_w^{-1}\int_{-\sigma_w/2}^{\sigma_w/2}dX\,\theta(X,\tau)$ and $\langle\theta(X,\tau')\rangle_{\tau}=(\omega/T)\int_{\tau}^{\tau+T}d\tau'\,\theta(X,\tau')$, respectively. Close to the drift instability threshold, $d\langle\theta\rangle_X/dt$ has slow dynamics on time, and therefore $\chi(\tau)\simeq\langle\chi(\tau')\rangle_{\tau}$. Introducing a new bifurcation parameter $\delta_D:=\gamma_o-\gamma_D$ characterising the distance from the point of drift instability, expanding $\chi(\tau)$ as $\chi=\langle\chi(\tau')\rangle_{\tau}+\delta_D\langle\chi(\tau')\rangle_{\tau}-\delta_D^2\langle\chi(\tau')\rangle_{\tau}^2+\ldots$, and averaging Eq.~\eqref{Eq:PolarForm} on space and time, one obtains
\begin{equation}
\label{Eq:Chi}
 \frac{d\chi}{d\tau}=\frac{8\omega C_0}{\langle v\rangle}\left(1+2\delta_D\chi-3\delta_D^2\chi^2\right), 
\end{equation}
where $C_0:=\langle(\rho k)^{-1}\partial_X\theta_v\partial_X\rho\rangle_{X, \tau}$ and $\langle v\rangle:=d\langle\theta(X,\tau)\rangle_X/dt$ is the average phase speed of the drifting pattern. Solving analytically Eq.~\eqref{Eq:Chi}, and assuming that near the drift instability the frequency of the drifting waves increase linearly with $\delta_D$ as $\omega(\delta_D)=\Omega_0\delta_D+\mathcal{O}(\delta_D^2)$, we get the following maximum average speed
\begin{equation}
    \label{Eq:AverageSpeed}
    \langle v\rangle_{\max}=\sqrt{\beta\left(\gamma_o-\gamma_D\right)},
\end{equation}
with $\beta:=32\Omega_0C_0/3$. For $\gamma_o$ close to $\gamma_D$, the pattern velocity increases as the square root of $\gamma_o$.

\subsection{Experimental evidence of the drift instabilities in a fluid experiment}

\begin{figure}
\begin{centering}
   \scalebox{0.22}{\includegraphics{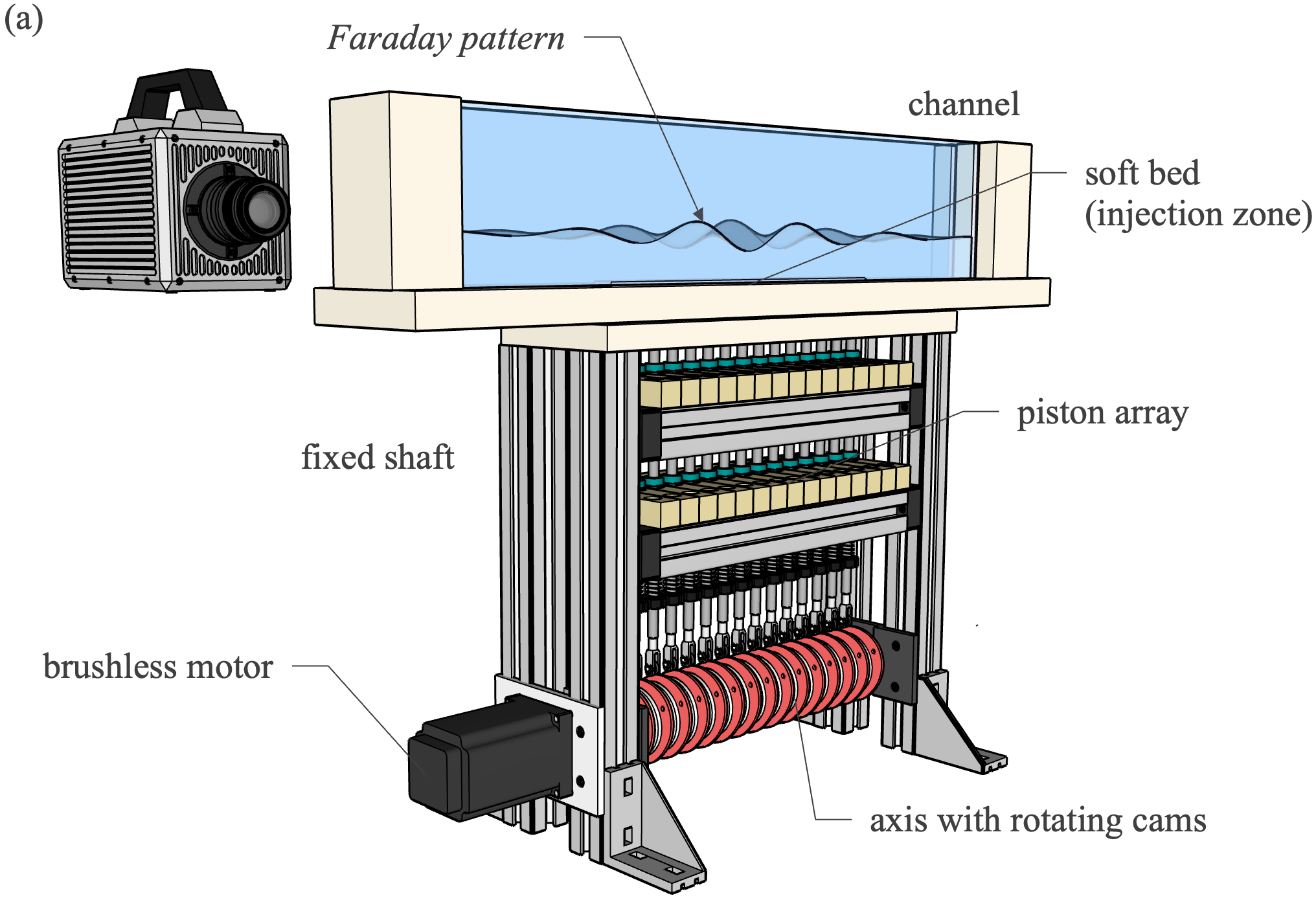}}\scalebox{0.265}{\includegraphics{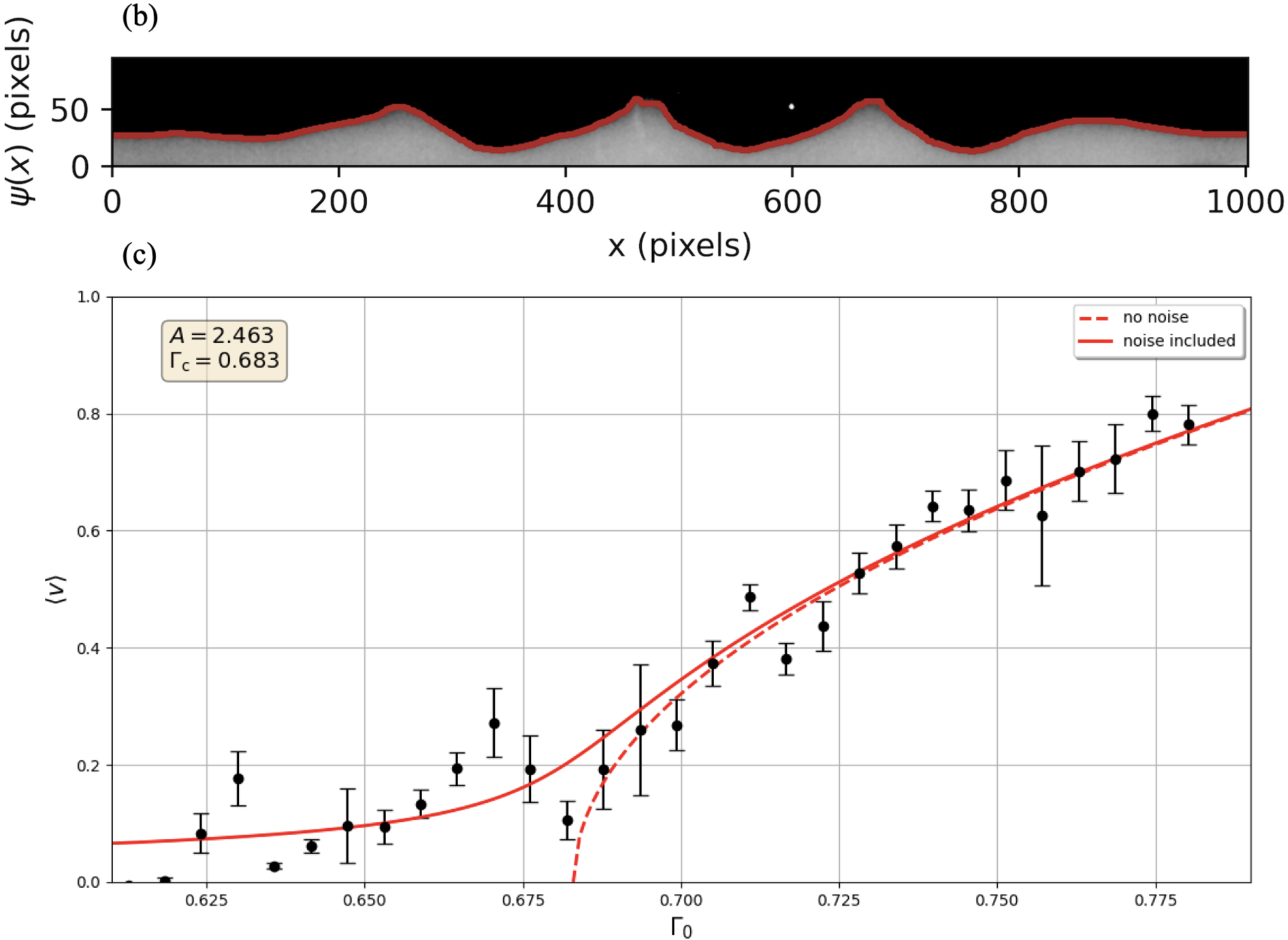}}\\
   \scalebox{0.4}{\includegraphics{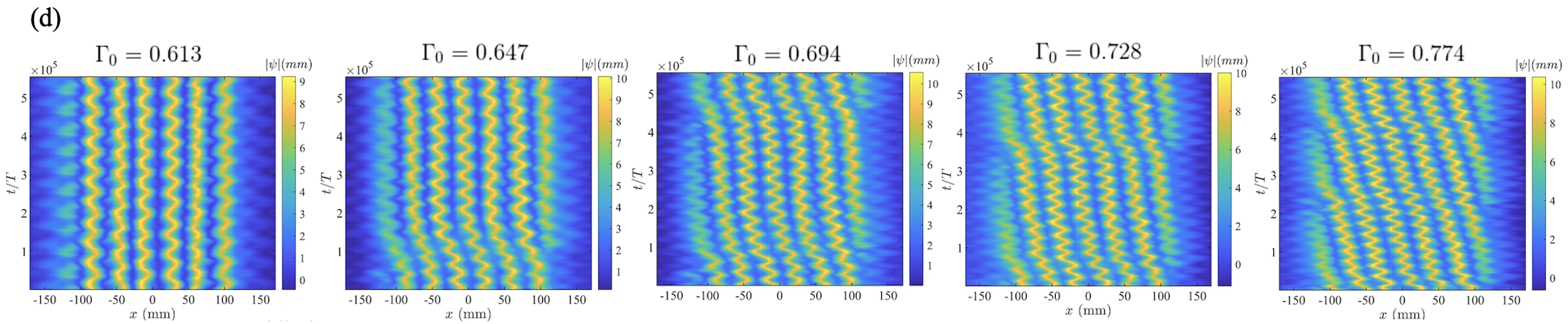}}
    \par\end{centering}
    \caption{\textbf{(a)} Experimental setup for generating Faraday patterns in the surface of fluid under localised vertical vibrations. \textbf{(b)} Typical detection of the free surface from the images received from the high-speed camera. \textbf{(c)} Experimental measurements of the mean speed $\langle v\rangle$ of the drifting patterns as a function of the acceleration amplitude $\Gamma_0$. The fit using theoretical predictions with and without noise effects is shown in solid and dashed red lines, respectively. \textbf{(d)} Typical behaviour of the drifting patterns in experiments under different acceleration amplitudes.} 
    \label{fig:04}
\end{figure}

Guided by the above theoretical study of drift instabilities, we have performed experiments of localised Faraday waves in a fluid subject to vertical and localised vibrations. The experimental setup is schematised in Fig.~\ref{fig:04}(a) and described in detail in Section \ref{Section:Methods}. Figure \ref{fig:04}(b) shows a typical detection of the free-surface elevation from the images obtained from the high-speed camera. It is known that the pdnlS equation~\eqref{Eq:01} governs the envelope of the free surface of a vertically vibrated fluid in the vicinity of Faraday instability \cite{Urra2019}. In particular, the pdnlS equation captures the water surface displacement of the transverse mode if the viscosity is small \cite{Miles1984}. The parameters in the dimensional pdnlS equation describing the envelope are determined by the characteristics of the fluid system, such as the injection frequency $\omega$ and the displacement amplitude of forcing $a_o$. These values are needed to compute $\gamma_o = \Gamma_o/4g$ and $\nu = (\omega^2/\omega_o^2 - 1)/2$, where $\Gamma_o = 4 \omega^2 a_o$ and $g$ is the acceleration of gravity.

We looked for the predicted drift instabilities in the fluid experiment shown in Fig.~\ref{fig:04}(a) starting from a fixed forcing frequency. We gradually increase the forcing amplitude by raising the displacement amplitude only. In Fig.~\ref{fig:03}(d), we show an experimental spatiotemporal diagram were the Faraday pattern exhibit absolute stability. Notice that the maximum wave amplitude exhibits a shift towards the left, in accordance with our previous observations in Fig.~\ref{fig:03}(a) and Fig.~\ref{fig:02}(a). Further increasing $\Gamma_0$, we were able to identify the unidirectional drift shown in Fig.~\ref{fig:03}(e). We notice by direct comparison between figures~\ref{fig:03}(b), \ref{fig:03}(e), \ref{fig:02}(d), and \ref{fig:02}(g) the correspondence between theory, pdnlS simulations, and experiments. Remarkably, the drift phenomenon is robust enough to be captured by the experiment despite the slight difference in scale given by the parameters of the equation. We also remark the appearance of a fast zig-zag kind of motion, whose wavelength seems to depend on the forcing amplitude. Such zig-zag movement appears above a certain instability threshold and will be studied elsewhere.

We identify a forcing amplitude dependence on the drift velocity. Figure~\ref{fig:04}(c) shows the experimental measurements of the mean drift velocity $\langle v\rangle$ for different values of $\Gamma_o$, whereas Fig.~\ref{fig:04}(d) summarises the different drifting states observed. The fit of the experimental speed measurements using Eq.~\eqref{Eq:AverageSpeed} is shown with dashed red line in Fig.~\ref{fig:04}(c). We notice that such a curve fits well the experimental observations for $\Gamma_0\gtrsim0.676$. However, considering the effects of additive noise, which are usually small and unavoidably present in experiments, the most probable value of the mean drift velocity follows the law \cite{Agez2013}
\begin{equation}
\label{Eq:08}
\langle v\rangle=A\sqrt{\frac{1}{2}\left(\Delta_D+\sqrt{\Delta_D^2+2\xi}\right)},    
\end{equation}
where $\Delta_D:=\Gamma_0-\Gamma_c$ and $\xi$ is the noise intensity. Figure \ref{fig:04}(c) shows the fitting curve using Eq.~\eqref{Eq:08} in solid red line. The outcome parameter values from the fitting procedure were $A=2.463$, $\Gamma_c=0.683$, and $\eta=0.01$. We observe a remarkable agreement between the experimental measurements and the fit using Eq.~\eqref{Eq:08}. Neglecting noise effects on the bifurcation, we have $\xi\to0$ in Eq.~\eqref{Eq:08}, and we recover our theoretical prediction on Eq.~\eqref{Eq:AverageSpeed}, which is valid for the deterministic case.

Further increasing $\Gamma_o$ above the drift instability threshold, we observe the complex convective state shown in Fig.~\ref{fig:03}(f). In this case, pattern dynamics exhibit branching phenomena, intermittence, and convection towards both possible directions. Simulations of the pdnlS equation~\eqref{Eq:01} exhibit spatiotemporal complexity near this combination of parameters, as shown in Fig.~\ref{fig:03}(c). The pdnlS approximation of the hydrodynamic equations describing the Faraday instability captures qualitatively well the observed phenomenon. However, the normal form of Eq.~\eqref{Eq:04} is valid for $\gamma_o\sim\gamma_o^{(0)}$ and cannot capture the complex convective states observed far above the instability threshold. Higher-order corrections would be needed to capture such spatiotemporal complexities in the amplitude equation.

\subsection{Impact of self-phase modulation terms on the nonlinear saturation of localised Faraday patterns}

Given the above theoretical and experimental evidence of drift instabilities in localised Faraday patterns, we wondered on the impact of such convection in the nonlinear saturation of the pattern. Following our previous calculations~\cite{Urra2019}, we perform a multiple-scale analysis in Eq.~\eqref{Eq:04} to investigate the role of the new SPM terms in the nonlinear saturation mechanism. We have obtained that SPM has no effects in the saturation of the fundamental mode close to its threshold of instability. Assuming that the spatial and temporal scale of perturbations on the wave amplitude is large, one obtains for $\gamma_o\sim\gamma_o^{(0)}$ that the evolution of the amplitude $D$ of the fundamental mode is governed by
\begin{equation}
 \label{Eq:05}
 \frac{d}{dt}D=\delta D-\frac{9}{2\sqrt{3}\mu}D^5,
\end{equation}
which is in full agreement with our previous results~\cite{Urra2019} obtained without SPM terms. Indeed, Eq.~\eqref{Eq:05} predicts the scaling law $D\propto\delta^{1/4}$, which has been previously verified in experiments~\cite{Urra2019} and gives the signature of a supercritical bifurcation. Figure~\ref{fig:01}(c) shows the maximum stationary amplitude $|C_{\max}|$ of the Faraday wave as a function of the pump strength $\gamma_o$. We show in solid line the prediction from Eq.~\eqref{Eq:05} and with stars the values obtained from the numerical simulations of the pdnlS Eq.~\eqref{Eq:01}. We conclude that the results of the multiple-scale analysis on the amplitude equation~\eqref{Eq:04} are in good agreement with the numerical simulations of the pdnlS equation~\eqref{Eq:01}.

\section{Discussion \label{Sec:Discussion}}

In a previous work~\cite{Urra2019} we have derived an amplitude equation describing the growth of nonlinear instabilities around the fundamental threshold $\gamma_o^{(0)}$, performing an appropriate expansion on the dispersion relation obtained from the WKBJ analysis. The resulting amplitude equation gives the first insight into how the nonlinear instability of the Gauss-Hermite modes gives rise to a saturated solution. However, until now such amplitude equation has not been formally derived using the standard tools from the weakly nonlinear analysis. 

The drift instabilities reported in this article are purely induced by the heterogeneity of the drive and weak asymmetries in the system, which are always present in real-life situations. Indeed, under a homogeneous drive, the amplitude of the Faraday pattern will be space-independent and the SPM terms causing the drift in Eq.~\eqref{Eq:04} vanishes. Notice that only the heterogeneous Weber-like term in Eq.~\eqref{Eq:04} vanishes after taking the limit $\sigma_i\to\infty$ (i.e.,  $\gamma(x)\to\gamma_o$). However, although the SPM terms are not directly coupled to the Gaussian width $\sigma_i$, they are \emph{indirectly} linked through the heterogeneous profile of $C(x,t)$ induced by the Gaussian. We believe this is the reason why such drift has not been observed in previous studies, which have been mainly focused on homogeneously driven systems. The role of heterogeneities is increasingly becoming an important factor in many nonlinear systems in different areas of physics \cite{Urra2019, Edri2020-1, Edri2020-2, Garcia-Nustes2021}. Our findings validate the use of non-local terms, such as those naturally appearing in Eq.~\eqref{Eq:04}, in other phenomenological models to capture heterogeneity-induced drift instabilities observed in simulations and experiments.

\section{Conclusions}

In summary, we have shown that heterogeneity can trigger drift instabilities in the evolving amplitude of localised Faraday patterns. Our analysis started from a heterogeneous version of the pdnlS equation as a model for parametrically induced waves. Under the assumption of a slowly varying Gaussian pump, we have derived the evolution equation governing the slow dynamics of the amplitude of the pattern. The resulting amplitude equation includes new non-local SPM terms that reveal complex dynamics in the underlying patterns. We have shown that above a secondary threshold of instability, such new terms trigger convection through a nonlinear symmetry breaking induced by nonlinear gradients. We performed a thorough search in an experiment of localised Faraday patterns in the free surface of fluid under localised parametric drive, and found the drift instability around the parameter values predicted by the amplitude equation. Numerical simulations of the pdnlS equation were also in remarkable agreement with theory and experiments. These results extends our general understanding of the dynamical effects of heterogeneities in nonlinear extended parametrical oscillators, and how pattern selection can be dramatically affected by the nature of the heterogeneous pump.

\section{Methods \label{Section:Methods}}

\subsection*{Normal forms}

The theory of normal forms gives a powerful method for the systematic construction of local, near-identity, and non-linear transformations to simplify the equations describing the dynamics of complex nonlinear problems near a bifurcation point \cite{Haragus2010}. We start writing Eq.~\eqref{Eq:01} in the form $\partial_t\boldsymbol\Psi =L\boldsymbol\Psi+\mathbf{N}$, where $\boldsymbol\Psi:=(\psi_R, \psi_I)$, $\psi:=\psi_R+i\psi_I$, $L$ is a linear operator, and $\mathbf{N}$ is a nonlinear vector. We perform our transformations around a Faraday wave solution with critical wavenumber $\kappa_c$ according to
\begin{equation}
\label{Eq:03}
\Psi=C(\eta, t)\left(\begin{array}{c}
        1 \\
       0
      \end{array}\right)e^{i\kappa_c x} + \sum_{n=1}^{\infty} \mathbf{W}^{[n]}+ c.c.,
\end{equation}
where $C(\eta, t)$ is a complex amplitude that varies slowly as a function of $\eta:=\epsilon^{1/2}x$, as described in Section~\ref{Sec:Results}, and the vector functions $\mathbf{W}^{[n]}\in\mathbb{C}^2,\,n\geq1$, are higher order corrections to be calculated. Assuming in Eq.~\eqref{Eq:01} the scaling laws $C\sim\epsilon^{1/4}$, $\delta\sim\epsilon$, $\partial_t\sim\epsilon$ and $\partial_x\sim\epsilon^{1/2}$, one can expand the pdnlS equation at different orders of $\epsilon^{1/4}$ and obtain a hierarchy of linear problems of the form
\begin{equation}
\label{Eq:Hierarchy}
    \mathcal{L}\mathbf{W}^{[n]}=\mathbf{b}_n,
\end{equation}
which can be solved for $\mathbf{W}^{[n]}$ only if $\mathbf{b}_n$ is in the image of the linear operator $\mathcal{L}$. According to the Fredholm alternative \cite{Pismen2006}, at least one solution for $\mathbf{W}^{[n]}$ exists if $\exists!|\mathbf{v}\rangle\in\ker\mathcal{L}^{\dagger}$ such that $\langle\mathbf{v}|\mathbf{b}_n\rangle=0$. Defining the inner product in the vector function space as
 \begin{equation}
 \label{Eq:InnerProd}
  \langle\mathbf{f}|\mathbf{g}\rangle=\kappa_c\int_{x_o}^{x_o+1/\kappa_c}dx\,(\mathbf{f}^*)^T\mathbf{g},\quad |\mathbf{f}\rangle, |\mathbf{g}\rangle\in\mathbb{C}^2,
 \end{equation}
the Fredholm alternative allows to obtain a solvability condition to calculate $\mathbf{W}^{[n]}$ at each order of $\varepsilon^{1/4}$. Our amplitude equation \eqref{Eq:04}
is the solvability condition obtained at order $\varepsilon^{5/4}$.

\subsection*{The method of multiple scales}

We have used the method of multiple scales to study the effects of SPM in the growth of the Faraday pattern near the fundamental threshold of instability. Such perturbation method allows to separate the time-scale dynamics of the different Gauss-Hermite modes controlling divergences in the perturbative developments \cite{Peyrard2004}.

Just above the fundamental threshold of instability, the fundamental Gauss-Hermite mode, $\mathcal{G}_0$, dominates the amplitude dynamics of the Faraday pattern; higher order modes are decaying modes \cite{Urra2019}. As the bifurcation parameter is increased, more Gauss-Hermite modes are excited. Each mode grows at their own growth rate/characteristic time. Assuming that the spatial and temporal scale of perturbations on the wave amplitude is large, the spatial derivative is re-scaled by a small parameter $\varepsilon$: $\partial_x\to\varepsilon^{1/2}\partial_x$, and we define slow time scales $T_i:=\varepsilon^i t$ (with $i=1,\,2,\,\ldots$). Writing $C$ as a perturbative development of the different time-scales
\begin{equation}
C(x,t):=\sum_{k=1}^{\infty}\varepsilon^{k/4}A_k(x, T_1, T_2, \ldots),
\end{equation}
and using a similar expansion in the bifurcation parameter, $\delta=\delta_o+\varepsilon\delta_1+\varepsilon^2\delta_2+\ldots$, 
one obtains at each order of $\varepsilon^{1/4}$ a hierarchy of problems similar to Eq.~\eqref{Eq:Hierarchy} for $A_k$,  $k\geq1$.

At order $\varepsilon^{1/4}$, one obtains that $A_1$ is governed by the (time-independent) Weber equation~\cite{Urra2019}, whose general solution is given by a linear combination of the Gauss-Hermite modes with time-dependent coefficients $D_n(T_1,\ldots)$, $n\geq0$. In particular, the saturation of the fundamental Gauss-Hermite mode  will be given by the time evolution of the first of such time-dependent coefficients, $D_o(T_1,\ldots)$.

At order $\varepsilon^{5/4}$, in the analogous hierarchical equation \eqref{Eq:Hierarchy} one obtains a vector $\mathbf{b}_5$ that contains SPM terms. Given that the set of all Gauss-Hermite modes $\{\mathcal{G}_n\}_{n=0}^{\infty}$ is a basis of the space, we write the functions $A_2$ and $A_3$ as the linear expansion
 \begin{equation}
  \label{AppB:Eq05}
  A_k=D_o^{(k)}(T_1)\mathcal{G}_0+\sum_{n=1}^{\infty}D_n^{(k)}(T_1)\mathcal{G}_n\quad,\quad k=2,3.
 \end{equation}
 However, near the threshold of instability of the fundamental mode, the modes $\mathcal{G}_{n}$ with $n\geq2$ are stable. Thus, the coefficients $D_n^{(k)}(T_1)$ for $n\geq2$ decays in time. Once the Faraday pattern is completely evolved, the maximum amplitude of the pattern is $C_{\max}\simeq D_o:=D$ and the functions $A_k$ ($k=1,2,3$) are even in space. Thus, using the Fredholm alternative, from symmetry arguments and the inner product~\eqref{Eq:InnerProd} one obtains that the contribution of the self-phase modulation terms vanish, and the solvability conditions leads to Eq.~\eqref{Eq:05}.

\subsection*{Numerical methods} Numerical simulations of the pdnlS equation \eqref{Eq:01} and our amplitude equation \eqref{Eq:04} where performed with no-flux boundary conditions, using a 401-point spatial grid with resolution $dx=0.25$. We used finite differences of second order of accuracy for the spatial derivatives, and the time integration was achieved using a fourth-order Runge-Kutta scheme with time-step $dt=0.0001$. To obtain the amplitude of the drifting localised Faraday patterns shown in Fig.~\ref{fig:03}, first we obtain numerically the pattern profile $\psi(x,t)$ from the numerical simulation of Eq.~\eqref{Eq:01}. Then, the complex amplitude $C$ is obtained by means of the Hilbert transform in space, $C(x,t)=(1/\pi)\int_{-L/2}^{L/2}\,d\chi \psi(\chi, t)/(x-\chi)$, which is computed at each time step of the numerical integration.

\subsection*{Experimental set-up}

The experimental setup is depicted in Fig.~\ref{fig:04}(a). We used a rectangular water channel of 15 mm long, 490 mm wide, and 100 mm deep whose bottom has a central rectangular soft silicone elastomer, type Smooth-on Dragon Skin 10 Medium. The Gaussian injection $\gamma(x)$ is done through this soft zone. For controlling the amplitude and frequency of the injection, we used a brushless servomotor with feedback (Model No. BLM-N23-50-1000-B) controlled with a DMC-40x0 motion controller from Galil Inc. The servomotor is coupled to an axis for transmitting the rotational motion to an evenly spaced 13-piston array ($\Delta x = 16$ mm), placed right underneath the soft injection zone. The pistons are placed in the axis with a set of rotatory cams for keeping them locked during operation. Over the pistons, there are aluminium bars connected to the soft zone. Each one of them has a tiny roller in contact with the piston, making the transmission of rotational motion to a linear one possible. The bars are pushed back towards the piston with springs. The setup resembles the mechanical transmission system of a music box, as shown in Fig.~\ref{fig:04}(a), and allows one to deform the bottom of the channel with spatial distribution.

For the experiment, the basin was filled up to 20 mm in height with water mixed with white ink for computational detection purposes. We used a Phantom VEO 440L camera controlled by the PCC 3.5 software provided along with the camera for the data acquisition. The resolution of the captures was 1280 x 252 pixels, and for every amplitude, the acquisition time was around 94 s at 400 frames per second. Images were saved in grayscale for easing the contour detection process.

Due to fluid inertia, a pre-processing of the images is necessary for greater amplitude experiments. A thin water layer is formed in the front of the basin, making the detection of the water-air interface more diffuse. To keep the focus only on the studied physical system, the brightness and gain of the images have been changed with the PCC software tools. The image processing has been made with a Python code using the OpenCV package for image manipulation. For localising the water-air interface, we used a Canny edge detector over the images previously processed with a Gaussian blur and Sobel filter. Iterating this process over time, we obtain the spatio-temporal water profiles $\eta(x, t)$ shown in Fig.~\ref{fig:03}(d-f) and \ref{fig:04}(d). As a post-process meant for direct comparison with simulations, a Hilbert transform was made along the time axis for every pixel in Matlab, obtaining the envelope of Faraday waves described by $\left |\psi(x, t) \right |$ in the pdnlS equation.

\begin{acknowledgments}
The authors thank Milena P\'aez-Silva for improvements in the experimental setup. R.R.A., S.C., and M.A.G-\~N. thanks the financial support of ECOS grant No. 200006. J.F.M. thanks ECOS-Sud project No. C15E06 for supporting a collaboration visit at Universit\'e de Lille, France, where part of this work was conceived. J.F.M. also thanks the financial support of ANID FONDECYT POSTDOCTORADO grant No. 3200499. M.A.G-\~N. acknowledges the financial support of ANID FONDECYT grant No. 1201434.
\end{acknowledgments}



\end{document}